\documentclass[longbibliography,slac_one]{revtex4}
\usepackage{graphicx}
\usepackage{fancyhdr}
\usepackage{subfigure}

\def \Bs {B_s^{0}}
\def \Bd {B^{0}}
\def \Bqb {\overline{B}{}_q^{0}}
\def \Bsb {\overline{B}{}_s^{0}}
\def \G {\Gamma}
\def \lf {\lambda}

\def \ch {\cosh \frac{\Delta \G_q t}{2}}
\def \cs {\cos(\Delta m_q t)}
\def \sh {\sinh \frac{\Delta \G_q t}{2}}
\def \sn {\sin(\Delta m_q t)}
\def \RE {{\rm Re}}
\def \IM {{\rm Im}}

\begin{document}

\title{$CP$ Violation in Other $\Bs$ Decays}

%

\author{L. Zhang (On behalf of LHCb Collaboration)}
\affiliation{Syracuse University, Syracuse, New York, 13244, USA}

\begin{abstract}
The recent experimental results of $CP$ violation in $\Bs$ decays other than in the $J/\psi\phi$ final state are discussed. Included are the resonant components and $\phi_s$ determination in $\Bs \to J/\psi \pi^+ \pi^-$, $CP$ asymmetries in $\Bs \to h^+ h^{(\prime)-}$ decays, and the $\Bs$ effective lifetimes in the $CP$-even state $K^+ K^-$ and the $CP$-odd state $J/\psi f_0(980)$.
\end{abstract}

\maketitle

\thispagestyle{fancy}


\section{Introduction}
The time evolution of the $B^0_q$-$\Bqb$ system is described by the
Schr\"{o}dinger equation
\begin{equation}
i\frac{\partial}{\partial
t}\left(\begin{array}{c}{|B^0_q}(t)\rangle\\{|\Bqb}(t)\rangle\end{array}\right)=
\left(\mbox{\boldmath $\rm M$}^q-\frac{i}{2}\mbox{\boldmath $\rm
\Gamma$}^q\right)
\left(\begin{array}{c}{|B^0_q}(t)\rangle\\{|\Bqb}(t)\rangle\end{array}\right),
\end{equation}
where the \mbox{\boldmath $\rm M$}$^q$ and \mbox{\boldmath $\rm \Gamma$}$^q$
matrices are Hermitian, and $CPT$ invariance requires
$M^q_{11}=M^q_{22}$ and $\Gamma^q_{11}=\Gamma^q_{22}$.
The off-diagonal elements, $M^q_{12}$ and $\Gamma^q_{12}$, of these matrices describe the off-shell (dispersive)
 and on-shell (absorptive) contributions
to $B^0_q$-$\Bqb$ mixing, respectively.

The mixing can be described by three physical quantities: $|M^q_{12}|$, $|\Gamma^q_{12}|$ and the relative phase $\phi^q_{12} = \arg\left(-\frac{M^q_{12}}{\Gamma^q_{12}}\right)$.  They are related to the experimental observables \citep{Lenz:2012mb} as: $\Delta M_q \simeq 2|M^q_{12}|$, $\Delta \Gamma_q \simeq 2|\Gamma^q_{12}|\cos\phi^q_{12}$, and $a^q_{\rm sl} \simeq \frac{\Delta \Gamma_q}{\Delta m_q}\tan \phi^q_{12}$, where $\Delta M_q$ is the mass difference between the heavy and the light mass eigenstates, $-\Delta \Gamma_q$ the width difference, and $a^q_{\rm sl}$ the semileptonic (or flavor specific) $CP$ asymmetry. The phase difference, denoted as $\phi_q$, between $\arg(M_{12}^q)$ and the decay phase in $b\to c\bar{c}s$ transitions is also an experimental observable. In the $B_s$ case, $\phi_s = -2\arg[V_{ts}V_{tb}^*/V_{cs}V_{cb}^*]$, is small and accurately predicted as $-0.0363^{+0.0016}_{-0.0015}$ rad~\citep{Charles:2011va}. New Physics \citep{Bobeth:2011st} in mixing could add new phases to $M_{12}^q$ or/and $\Gamma_{12}^q$ and modify $\phi^q_{12}$ or/and $\phi_q$ from their Standard Model (SM) predictions.

\section{Formalism}
The time dependent decay rates for initial $B^0_q$ or $\Bqb$ decays to a $CP$ eigenstate ($f_{\rm CP}$) are:
\begin{eqnarray}\label{Eq-t}
\Gamma(B_q^0(t)\to f_{\rm CP})= {\cal N} e^{-\G_q t}\left\{\frac{1+|\lf|^2}{2}\ch  + \frac{1-|\lf|^2}{2}\cs
-\RE(\lf)\sh  -  \IM(\lf)\sn\right\},\nonumber\\
\Gamma(\overline{B}_q^0(t)\to f_{\rm CP})= {\cal N} e^{-\G_q t}  \left\{\frac{1+|\lf|^2}{2}\ch -  \frac{1-|\lf|^2}{2}\cs
-\RE(\lf)\sh + \IM(\lf)\sn\right\},
\end{eqnarray}
where $\lambda = (q/p)(\bar{A}_f/A_f)$, $\G_q$ is average width of two mass eigenstates, $A_f$ ($\bar{A}_f$) is the amplitude of $B^0_q$ ($\Bqb$) decay, and $|p/q|=1$ is used assuming no $CP$ violation in the mixing.

The $CP$ asymmetry is\begin{equation}
A_{f_{CP}}(t)\equiv \frac{\Gamma(\overline{B}_q^0(t)\to f_{CP})-\Gamma(B_q^0(t)\to f_{CP})}{\Gamma(\overline{B}_q^0(t)\to f_{CP})+\Gamma(B_q^0(t)\to f_{CP})}
=\frac{\displaystyle {A^{\rm dir}}\cs + {A^{\rm mix}} \sn}{\displaystyle \ch + {A^{\Delta \G}}\sh}.
\end{equation}
There are three $CP$ variables but only two are independent: direct $CP$ asymmetry $A^{\rm dir}=\frac{\displaystyle |\lf|^2-1}{\displaystyle |\lf|^2+1}$,
mixing induced $CP$ asymmetry $A^{\rm mix}=\frac{\displaystyle 2\IM \lf}{\displaystyle |\lf|^2+1}$ and
$A^{\Delta \G}=-\frac{\displaystyle 2\RE \lf}{\displaystyle |\lf|^2+1}$ satisfying $(A^{\rm dir})^2+(A^{\rm mix})^2+(A^{\Delta \G})^2=1$.

\section{Resonant components and $\phi_s$ determination in $\Bs \to J/\psi \pi^+ \pi^-$}
Motivated by a predication in Ref. \citep{Stone:2008ak}, the LHCb collaboration made the first observation of $\Bs \to J/\psi f_0(980), f_0(980)\to \pi^+\pi^-$ \citep{Aaij:2011fx}, which was subsequently confirmed by others \citep{Li:2011pg,Aaltonen:2011nk,Abazov:2011hv}. This mode is a $CP$-odd eigenstate and can be used to determine $\phi_s$ without the need for an angular analysis, as is required in the $J/\psi \phi$ final state \citep{LHCb:2011aa,Abazov:2011ry,CDF:2011af}. With 0.4~fb$^{-1}$ data, LHCb used the candidates within $\pm90$ MeV of $f_0(980)$ mass and measured $\phi_s=-0.44\pm0.44\pm0.02$ rad \citep{LHCb:2011ab}. Whenever two uncertainties are given, the first is statistical and the second systematic. However, the used events are only about half of $J/\psi \pi^+ \pi^-$ signal.

\begin{figure*}[h!t!]
\centering
\includegraphics*[width=0.7\textwidth, bb= 0 539 530 843]{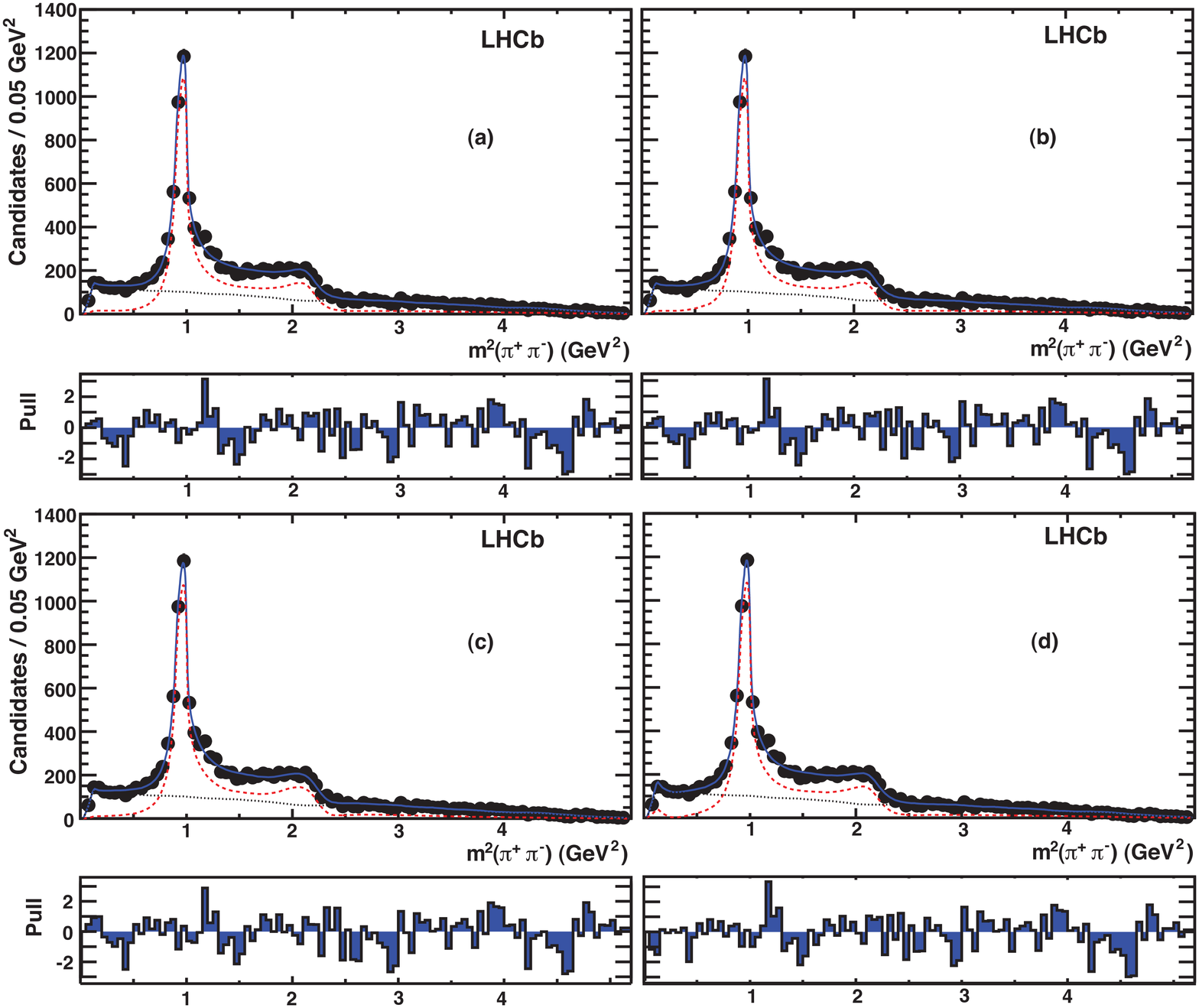}
\caption{Dalitz fit project of $m^2(\pi^+\pi^-)$ from the best fit. The points with error bars are data, the signal fit is shown with a (red) dashed line, the background with a (black) dotted line, and the (blue) solid line represents the total.} \label{DP1}
\end{figure*}

To optimize $J/\psi \pi^+ \pi^-$ usefulness, we need to understand the $CP$ content of this final state. With 1.0~fb$^{-1}$ data, LHCb \citep{LHCb:2012ae} preformed a modified Dalitz-plot analysis which fits to the $m^2(\pi^+\pi^-)$, $m^2(J/\psi \pi^+)$ and $J/\psi\to \mu^+\mu^-$ helicity angle ($\theta_{J/\psi}$) distributions after integrating the angle between $J/\psi$ and $\pi^+\pi^-$ decay planes. Different from the classical ``Dalitz-plot'' analysis, the vector $J/\psi$ particle has 3 helicity amplitudes that must be considered. The projection of $m^2(\pi^+\pi^-)$ overlayed with the best fit is shown in Fig. \ref{DP1}. The components and fractions of the best fit are given in Table~\ref{tab:ff1}. The fraction of $J/\psi \rho$ is measured less than 1.6\% at 95\% confidence level (CL); the $J/\psi \rho$ final state only can be present in higher order processes. The final states listed in Table \ref{tab:ff1} are all $CP$-odd states, expect for $f_2(1270)$ with helicity $|\Lambda| = 1$ which has mixed-$CP$. Combining with this and $\rho$, the fraction of $CP$-even states is less than 2.3\% at 95\% CL. So the whole mode is dominated by $CP$-odd state, and can be used for $\phi_s$ measurement without need of angular analysis. The relative branching ratio between $\Bs\to J/\psi \pi^+\pi^-$ and $J/\psi\phi$ is measured as (19.79$\pm$0.47$\pm0.52$)\% \citep{LHCb:2012ae}.

\begin{table}[h!]
\begin{center}
\caption{Resonance fractions in $\Bs\rightarrow J/\psi \pi^+\pi^-$ over the full mass range \citep{LHCb:2012ae}. The final-state helicity of the D-wave is denoted by $\Lambda$. Only statistical uncertainties are quoted.}\label{tab:ff1}
\begin{tabular}{lcc}
\hline
~~~Resonance &  Normalized fraction (\%)\\
\hline
$f_0(980)$ &$69.7\pm2.3$ \\
$f_0(1370)$ &$21.2\pm2.7$\\
non-resonant $\pi^+\pi^-$ & ~\,$8.4\pm1.5$\\
$f_2(1270)$, $\Lambda=0$ &~\,$0.49\pm0.16$\\
$f_2(1270)$, $|\Lambda| = 1$ &~\,$0.21\pm0.65$\\
\hline
\end{tabular}
\end{center}
\end{table}

The $\Bs\to J/\psi \pi^+\pi^-$ decay complements $J/\psi \phi$ in determining $\phi_s$. The event yield is about 40\% of $J/\psi \phi (\phi \to K^+ K^-)$. In $\Bs$ decay to a $CP$-odd final state via $b\to c\bar{c}s$ transition, the time dependent decay rates are given in Eq.~(\ref{Eq-t}) with $\lambda = -e^{-\phi_s}$. We also need to take into account experimental effects, such as acceptance in decay time, time resolution, and dilution due to wrong flavor tagging. The decay time resolution in LHCb is about 40 fs thanks to the excellent vertex detector and large momentum of $\Bs$.

\begin{figure*}[h!t!]
\centering
\includegraphics[width=0.44\textwidth]{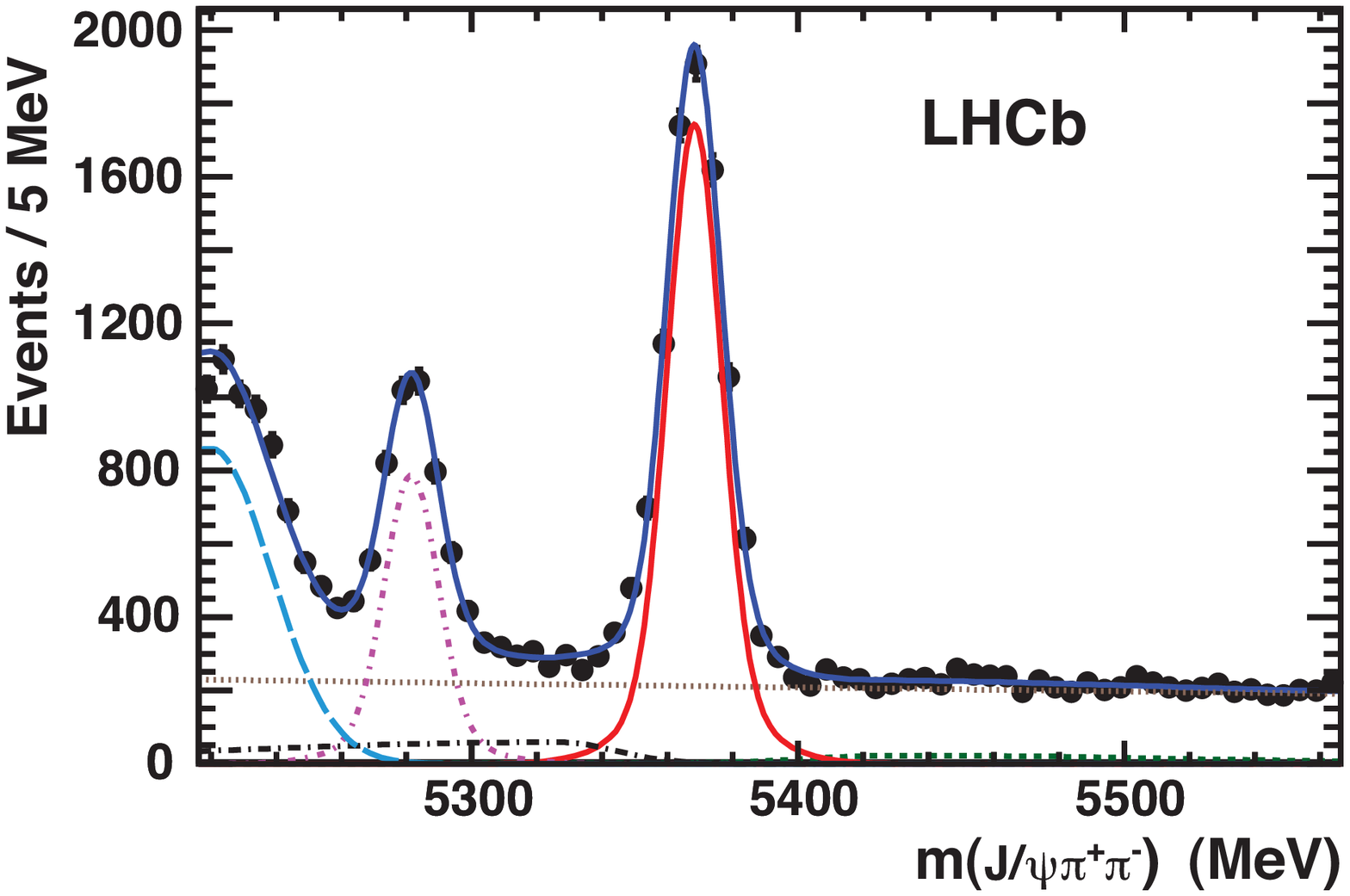}%
\includegraphics[width=0.46\textwidth]{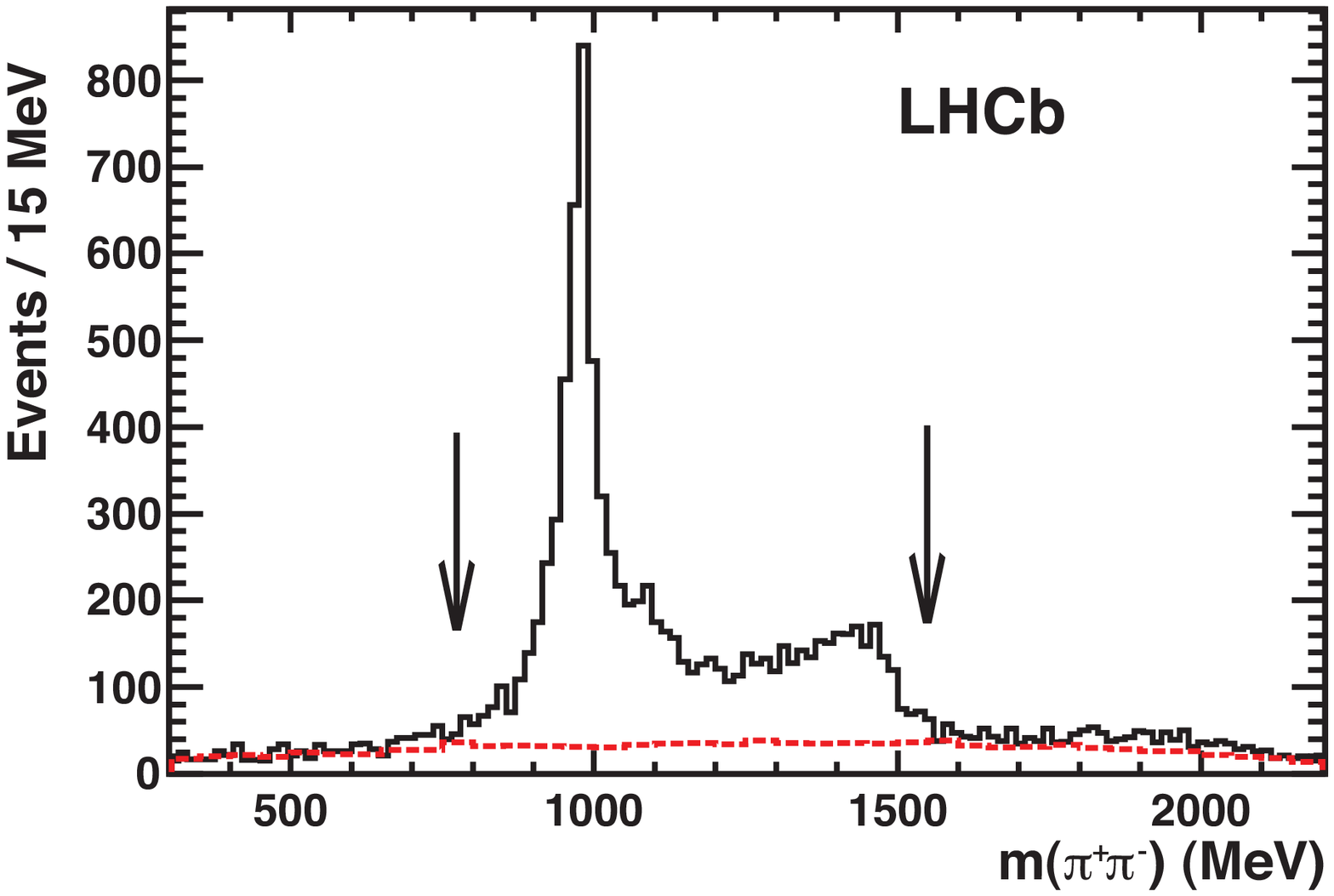}
\caption{Left: The $m(J/\psi \pi^+\pi^-)$ distribution from the candidates with $m(\pi^+\pi^-)\in [775,1550]$ MeV, shown also the $\Bs$ signal as the red solid line and other various background components. Right: The $m(\pi^+\pi^-)$ from the candidates with $\pm20$ MeV around $\Bs$ mass peak. The arrows indicate the region used for $\phi_s$ measurement and the red dash line shows the background.} \label{jpsipipi}
\end{figure*}

LHCb updated the $\phi_s$ measurement using an sample of 1 fb$^{-1}$ data \citep{LHCb:2012ad}. The $m(J/\psi \pi^+\pi^-)$ and $m(\pi^+\pi^-)$ distributions are shown in Fig. \ref{jpsipipi} after applying a Boosted Decision Tree selection rather than a cut-based selection used in Fig. \ref{DP1} . With about 7400 $\Bs$ signal candidates, LHCb finds $\phi_s = -0.019^{+0.173+0.004}_{-0.174-0.003}$ rad, consistent with the SM expectation~\citep{Charles:2011va}. The largest systematic uncertainty arises from allowing direct $CP$ violation. Here $\phi_s$ changes by $-0.0020$ from the default value and $|\lambda|=0.89\pm0.13$ consistent with no direct $CP$ violation. The systematic uncertainty due to a possible $CP$-even component is $-0.0008$. The measured $\phi_s$ is consistent with the LHCb preliminary result $-0.001\pm0.101\pm0.027$ rad using $\Bs\to J/\psi \phi$ \citep{LHCb-CONF-2012-002}. Combining the two LHCb measurements gives $\phi_s=-0.002\pm0.083\pm0.027$ rad \citep{LHCb-CONF-2012-002}.

Other possible modes for measuring $\phi_s$ are
 \begin{itemize}
   \item $\Bs\to J/\psi f_2^{\prime}(1525)$ which was first observed by LHCb \citep{Aaij:2011ac}. This mode could be useful but it needs to include additional D-wave in transversity amplitudes. Sizeable S-wave over the entire $m(K^+K^-)$ in $J/\psi K^+K^-$ is also seen \citep{Aaij:2011ac}.
   \item $\Bs\to J/\psi \eta^{(\prime)}$ which were first observed by Belle \citep{Belle:2012aa}. They are $CP$-even states with large branching fraction, but neutral detection is difficult for experiments in a hadron collider.
   \item $\Bs\to \psi(2S) \phi, \psi(2S)\to \mu^+\mu^-$ and $\Bs \to D_s^+ D_s^-$. The event yields are of order of 5-10\% of $(J/\psi \to \mu^+\mu^-)(\phi\to K^+ K^-)$.
 \end{itemize}

\section{$CP$ asymmetries in $\Bs \to h^+ h^{(\prime)-}$ decays}
Two-body charmless $\Bs$ decays have significant contribution of penguin diagrams, providing an entry point for New Physics. 
Under U-spin symmetry, where $s$ quarks are changed to $d$ quarks, $\Bs \to K^+ K^-$ is analogous to $B^0 \to K^+\pi^-$ and $\Bs \to K^-\pi^+$ to $B^0\to \pi^+ \pi^-$. Specifically it turns out that $A_{KK}^{\rm dir}$ in $\Bs$ should be equal to the direct $CP$ asymmetry in $B^0 \to K^+\pi^-$, in the limit of exact U-spin symmetry and neglecting certain diagrams with annihilation topologies contributing to the decay amplitude \citep{Fleischer:1999pa}.

\begin{figure*}[t]
\centering
\includegraphics[width=0.9\textwidth]{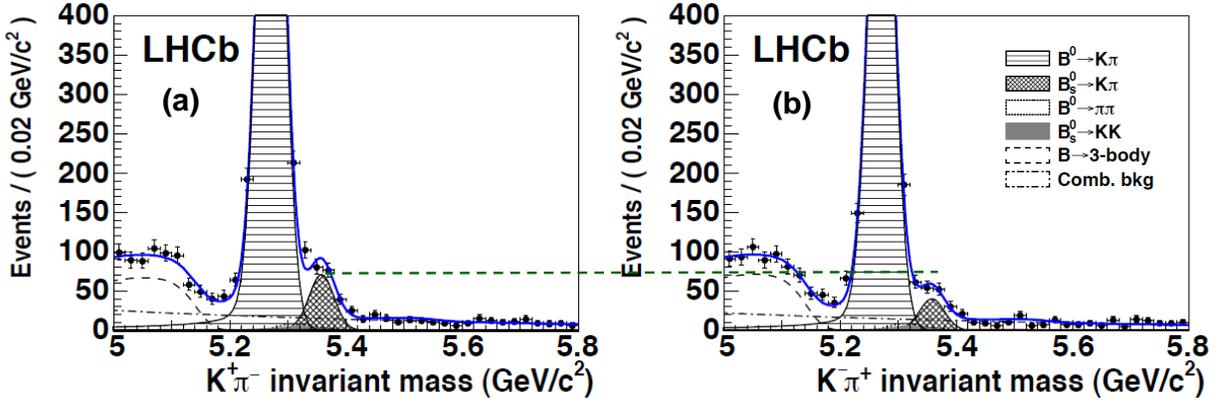}
\caption{Invariant $K\pi$ mass spectra for (a) $K^+\pi^-$ and (b) $K^-\pi^+$ combinations.} \label{acp}
\end{figure*}

The direct $CP$ asymmetry in the flavor-specific decay $\Bs \to (f = K^- \pi^+)$,
\begin{equation}
A_{\rm CP}(\Bs \to K\pi) = \frac{\displaystyle |\overline{A}_{\bar{f}}|^2-|A_f|^2}{\displaystyle |\overline{A}_{\bar{f}}|^2+|A_f|^2},
\end{equation}
is measured by LHCb using 0.35 fb$^{-1}$ data, where an untagged time-integrated method is used. The measured charge asymmetry after correcting detection and production biases is equal to $A_{\rm CP} +  {\cal O}(a^s_{\rm sl})$. The semileptonic asymmetry is measured as $a^s_{\rm sl}=(-1.81\pm1.06)\%$ by the D0 Collaboration~\citep{Abazov:2011yk}, where the uncertainty is sum of statistical and systematic. The LHCb Collaboration recently measured $a^s_{\rm sl}=(-0.24\pm0.54\pm0.33)\%$ \citep{LHCb-CONF-2012-022}. Thus ${\cal O}(a^s_{\rm sl})$ is negligible compared to the measured $CP$ asymmetry shown later. The $K^+\pi^-$ and $K^-\pi^+$ mass spectra for the candidates are shown in Fig. \ref{acp}. A Clear difference in the yields of two peaks at the $\Bs$ mass is seen.
LHCb finds $A_{\rm CP}(\Bs \to K\pi) = 0.27\pm0.08\pm0.02$ \citep{Aaij:2012qe}. It is 3.3$\sigma$ from zero, providing the first evidence for $CP$ violation in the decays of $\Bs$ mesons. The result for $A_{\rm CP}(\Bs \to K\pi)$ is consistent with the only previous measurement from CDF of $0.39\pm0.15\pm0.08$ \citep{Aaltonen:2011qt}.

\begin{figure*}[h!t!]
\centering
\includegraphics[width=0.45\textwidth]{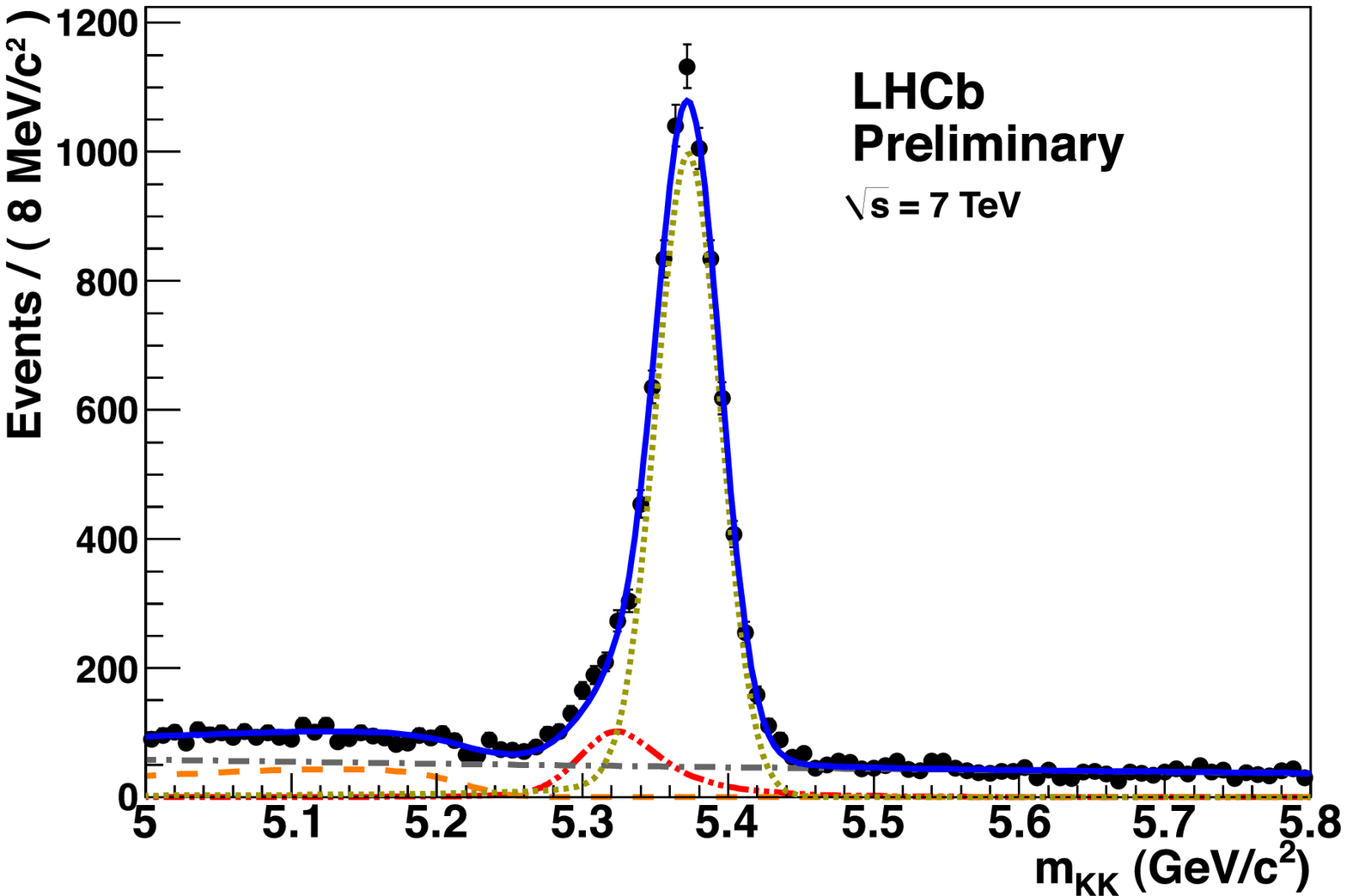}%
\includegraphics[width=0.44\textwidth]{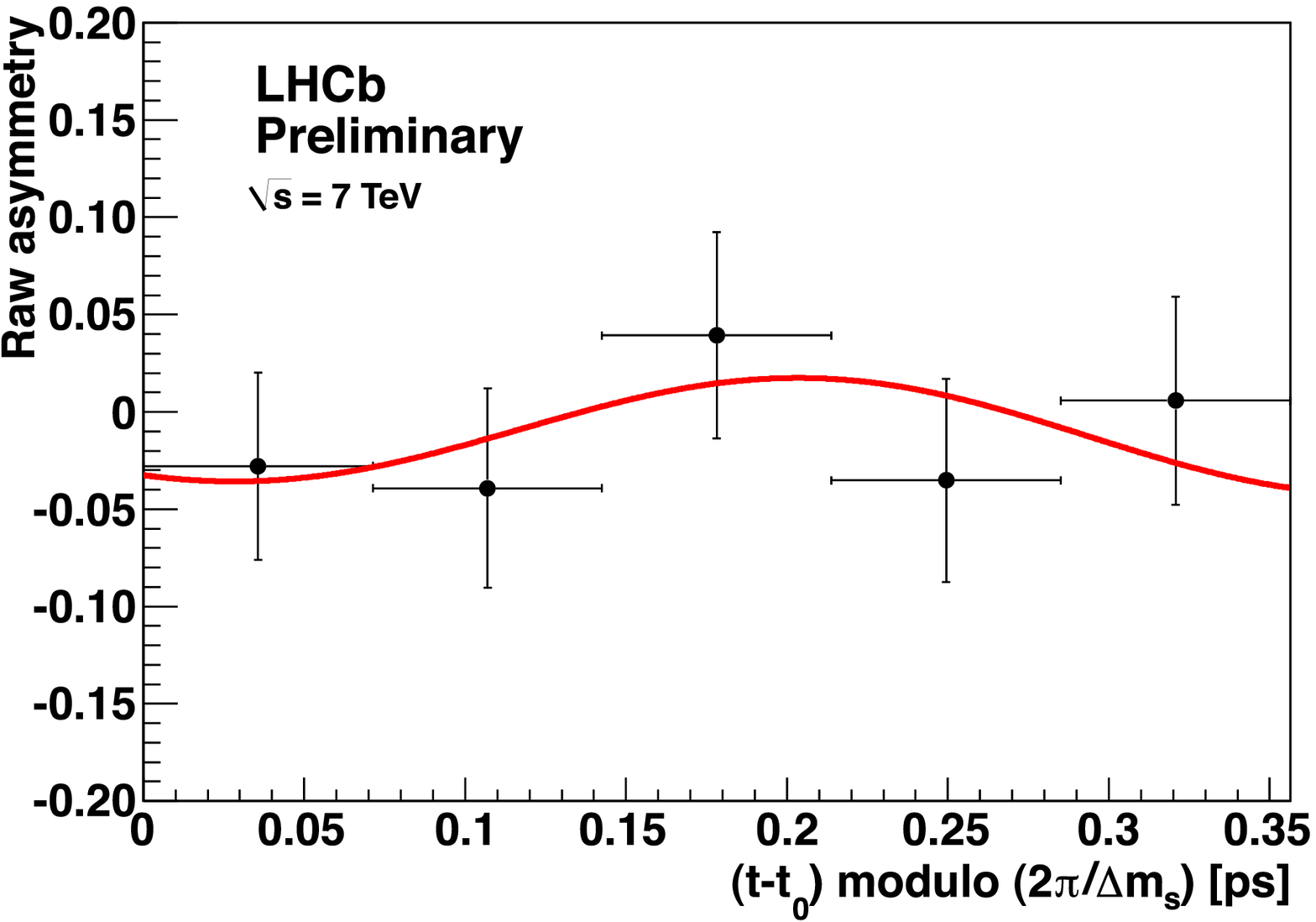}
\caption{Left: Invariant mass projection for the $\Bs\to K^+ K^-$. The data are described by the overall fit result (solid blue line) which is the sum
of: $\Bs \to K^+ K^-$ signal (dotted dark yellow line); $\Bd\to K\pi$ cross-feed background
(dashed double-dotted red line); partially reconstructed three-body (dashed orange line)
and combinatorial (dashed dotted grey line) backgrounds. Right: Time-dependent raw asymmetry in the $\Bs\to K^+K^-$ signal mass region with the
result of the fit superimposed. The offset $t_0 = 0.95$ ps corresponds to the selection requirement
where the decay time acceptance starts.} \label{acpkk}
\end{figure*}

LHCb \citep{LHCb-CONF-2012-007} made the first measurement of $CP$ asymmetry in $\Bs \to K^+K^-$. It's a time-dependent flavor-tagged  analysis. The large statistical sample $\Bd \to K^- \pi^+$ is used to measure flavor tagging efficiency and mistag probability that are then constrained in the fit to $\Bs \to K^+K^-$. The mass distribution of selected candidates is shown in Fig. \ref{acpkk} (left); fit obtains $7155\pm97$ signal candidates. The measured $CP$ asymmetries (preliminary) are
\begin{eqnarray}
A_{KK}^{\rm dir}&=&0.02\pm0.18\pm0.04,\nonumber\\
A_{KK}^{\rm mix}&=&0.17\pm0.18\pm0.05.
\end{eqnarray}
The raw asymmetry is shown in Fig. \ref{acpkk} (right).

\section{Effective lifetimes of $\Bs$ from $CP$-eigenstate final states}
The untagged decay time distribution summing $\Bs$ and $\Bsb$ is
\begin{equation}
\Gamma(t) \propto (1-A^{\Delta \Gamma_s})e^{-\Gamma_Lt}+(1+A^{\Delta \Gamma_s})e^{-\Gamma_Ht},
\end{equation}
where $A^{\Delta \Gamma_s}=-\frac{\displaystyle 2\RE \lf}{\displaystyle |\lf|^2+1}$, as in Eq.~(\ref{Eq-t}), and $\tau_H$ and $\tau_L$ are the lifetimes of heavy and light mass eigenstates. The $CP$-odd state $J/\psi f_0(980)$ has $A^{\Delta \Gamma_s}=\cos\phi_s=1$ and $CP$-even state $K^+K^-$ has $A^{\Delta \Gamma_s}\approx-1$ (the small difference from unity is discussed in Ref. \citep{Fleischer:2011cw}). Therefore, their effective lifetimes are approximately equal to the lifetimes of heavy and light mass eigenstates, respectively.

The effective lifetime in $J/\psi f_0(980)$ final state is measured to be $1.700 \pm 0.040 \pm 0.026$ ps by LHCb \citep{Aaij:2012nt} and $1.70^{+0.12}_{-0.11}$$\pm0.03$ ps by CDF \citep{Aaltonen:2011nk}. The most precise measurement on the effective lifetime in $K^+K^-$ final state comes from LHCb. The measured value is $1.455\pm0.046 \pm0.006$ ps \citep{:2012ns}. Those measurements can be used to improve the determination of $\Gamma_s$ and $\Delta \Gamma_s$ using the method carried out by HFAG ~\citep{Amhis:2012bh}.

\section{Summary}
In summary, the latest world best measurements on $CP$ violation in other $\Bs$ decays are discussed. They are mostly provided by LHCb collaboration thanks to the larger statics and better detector performance.
\bigskip 
\begin{acknowledgments}
Work supported by U.S. National Science Foundation.
\end{acknowledgments}

\bigskip 
\newpage

\end{document}